\documentclass[aps,prc,twocolumn,superscriptaddress,showpacs,showkeys]{revtex4-1}
\usepackage{graphicx}
\begin{document}
\title{New limits on heavy sterile neutrino mixing in ${^{8}\rm{B}}$-decay obtained with the Borexino detector}

\author{G.~Bellini$^h$, J.~Benziger$^k$, D.~Bick$^s$, G.~Bonfini$^e$, D.~Bravo$^q$, M.~Buizza Avanzini$^h$,
 B.~Caccianiga$^h$, L.~Cadonati$^p$, F.~Calaprice$^l$, P.~Cavalcante$^e$, A.~Chavarria$^l$, A.~Chepurnov$^r$,
 D.~D{\textquoteright}Angelo$^h$,
S.~Davini$^t$, A.~Derbin$^m$, I.~Drachnev$^m$, A.~Empl$^t$, A.~Etenko$^g$, K.~Fomenko$^{b,e}$, D.~Franco$^{a}$, C.~Galbiati$^l$, S.~Gazzana$^e$,
C.~Ghiano$^a$, M.~Giammarchi$^h$, M.~G\"{o}ger-Neff$^n$, A.~Goretti$^l$, L.~Grandi$^l$, C.~Hagner$^s$, E.~Hungerford$^t$, Aldo Ianni$^e$, Andrea
Ianni$^l$,  V.~Kobychev$^f$, D.~Korablev$^b$, G.~Korga$^t$, D.~Kryn$^a$, M.~Laubenstein$^e$, T.~Lewke$^n$, E.~Litvinovich$^g$, B.~Loer$^l$,
F.~Lombardi$^e$, P.~Lombardi$^h$, L.~Ludhova$^h$, G.~Lukyanchenko$^g$, I.~Machulin$^g$, S.~Manecki$^q$, W.~Maneschg$^i$, G.~Manuzio$^c$,
Q.~Meindl$^n$, E.~Meroni$^h$, L.~Miramonti$^h$, M.~Misiaszek$^d$, P.~Mosteiro$^l$, V.~Muratova$^m$, L.~Oberauer$^n$, M.~Obolensky$^a$, F.~Ortica$^j$,
K.~Otis$^p$, M.~Pallavicini$^c$, L.~Papp$^{q}$, L.~Perasso$^c$, S.~Perasso$^c$, A.~Pocar$^p$, G.~Ranucci$^h$, A.~Razeto$^e$, A.~Re$^h$,
A.~Romani$^j$, N.~Rossi$^e$, R.~Saldanha$^l$, C.~Salvo$^c$, S.~Sch\"onert$^{n}$, H.~Simgen$^i$, M.~Skorokhvatov$^g$, O.~Smirnov$^b$, A.~Sotnikov$^b$,
S.~Sukhotin$^g$, Y.~Suvorov$^{u,g}$, R.~Tartaglia$^e$, G.~Testera$^c$, D.~Vignaud$^a$, R.B.~Vogelaar$^q$, F.~von Feilitzsch$^n$, J.~Winter$^n$,
M.~Wojcik$^d$, A.~Wright$^l$, M.~Wurm$^s$, J.~Xu$^l$, O.~Zaimidoroga$^b$, S.~Zavatarelli$^c$,and G.~Zuzel$^{d}$
\\(Borexino Collaboration) \\}

\affiliation{ a) APC, Univ. Paris Diderot, CNRS/IN2P3, CEA/Irfu, Obs. de Paris, Sorbonne Paris Cit\'e, France\\
b) Joint Institute for Nuclear Research, Dubna 141980, Russia\\
c) Dipartimento di Fisica, Universit\`{a} e INFN, Genova 16146, Italy\\
d) M. Smoluchowski Institute of Physics, Jagiellonian University, Crakow, 30059, Poland\\
e) INFN Laboratori Nazionali del Gran Sasso, Assergi 67010, Italy\\
f) Institute for Nuclear Research, Kiev 06380, Ukraine\\
g) NRC Kurchatov Institute, Moscow 123182, Russia\\
h) Dipartimento di Fisica, Universit\`{a} degli Studi e INFN,
Milano 20133, Italy\\ i) Max-Plank-Institut f\"{u}r Kernphysik, Heidelberg 69029, Germany\\ j) Dipartimento di Chimica, Universit\`{a} e INFN, Perugia 06123, Italy\\
k) Chemical Engineering Department, Princeton University, Princeton, NJ 08544, USA\\ l) Physics Department, Princeton
University, Princeton, NJ 08544, USA\\ m) St. Petersburg Nuclear Physics Institute, Gatchina 188350, Russia\\ n) Physik
Department, Technische Universit\"{a}t M\"{u}nchen, Garching 85747, Germany\\ p) Physics Department, University of
Massachusetts, Amherst MA 01003, USA\\ q) Physics Department, Virginia Polytechnic Institute and State University,
Blacksburg, VA 24061, USA\\ r) Lomonosov Moscow State University Skobeltsyn Institute of  Nuclear Physics, Moscow
119234, Russia\\ s) Institut f\"ur Experimentalphysik, Universit\"at Hamburg, Germany\\ t) Department of Physics,
University of Houston, Houston, TX 77204, USA\\ u) Physics ans Astronomy Department, University of California Los
Angeles (UCLA), Los Angeles, CA 90095, USA}

\begin{abstract}
If heavy neutrinos with mass $m_{\nu_{H}}\geq$2$ m_e $ are produced in the Sun via the decay ${^8\rm{B}} \rightarrow
{^8\rm{Be}} + e^+ + \nu_H$ in a side branch of pp-chain, they would undergo the observable  decay into an electron, a
positron and a light neutrino $\nu_{H}\rightarrow\nu_{L}+e^++e^-$. In the present work Borexino data are used to set a
bound on the existence of such decays. We constrain the mixing of a heavy neutrino with mass 1.5 MeV $\leq m_{\nu_{H}}
\le$ 14 MeV to be $|U_{eH}|^2\leq (10^{-3}-4\times10^{-6})$ respectively.  These are tighter limits on the mixing
parameters than obtained in previous experiments at nuclear reactors and accelerators.
\end{abstract}



\maketitle

\section{Introduction}
Neutrino flavor oscillations show that neutrinos are massive. In turn, heavier neutrinos can decay into lighter ones \cite{Shr81}-\cite{Gor07}.
Within the framework of an extended standard model (SM), the simplest detectable decay modes are the radiative decay
$\nu_{H}\rightarrow\nu_{L}+\gamma$ and the decay into an electron-positron pair plus a light neutrino:
\begin{equation}
\nu_{H}\rightarrow\nu_{L}+e^++e^- ,
\end{equation}
resulting from the $W$ and $Z$ exchange diagrams shown in fig.~\ref{Figure:Decay_PC} (b and c) and possible only for $m_{\nu_{H}}\geq 2m_e$. The $Z$
exchange also results in the invisible decay mode $\nu_{H}\rightarrow 3\nu$, (fig.~\ref{Figure:Decay_PC}, d). Since neutrino oscillations are well
described in terms of the three known light neutrinos, a heavy neutrino would necessarily need to be a nonstandard new particle.



Many extensions of the SM predict the existence of one or more sterile neutrinos, which appear as  singlet fermions in $\nu\rm{MSM}$ neutrino in
minimal standard model \cite{Gor07,Gni13}, mirror neutrinos \cite{Ber03}, goldstinos in SUSY \cite{Chu96}, modulinos of the superstring theories
\cite{Ben97}, or bulk fermions linked to the existence of extra dimensions \cite{Hol02, Cir05, Moh06}. In general, massive sterile neutrinos may have
arbitrary mass and mix with the active flavors.

In the heavy neutrino rest frame, the decay rate for $(\nu_{H} \rightarrow \nu_{L}e^+e^-)$ due to $W$ exchange was obtained in
\cite{Shr81}-\cite{Pal82}. The total decay rate for all three modes shown in fig.~\ref{Figure:Decay_PC} including $Z$ exchange and invisible decay
$(\nu_{H} \rightarrow \nu_{e} \nu_{e,\mu,\tau} \widetilde{\nu}_{e,\mu,\tau}$) is \cite{Gor07}:


\begin{equation}
\Gamma_{c.m.}^{tot}\cong\frac{{G_F}^2}{192\pi^{3}}{m_{{\nu}_H}^5}|U_{eH}|^2(1+h\left[\frac{{m_e}^2}{m_{{\nu}_H}^2}\right])
\label{tau_cm}
\end{equation}
where $U_{eH}$ is the mixing parameter of the heavy neutrino to the electron flavor, ${G_F}$ the weak coupling constant
and $h[{m_e}^2/m_{{\nu}_H}^2]$ the phase-space factor calculated in \cite{Gor07}.  The
$\nu_{H}\rightarrow\nu_{e}e^+e^-$ decay rate $\Gamma_{c.m.}^{e^+e^-}$ due to $W$- and $Z$-exchange
(fig.~\ref{Figure:Decay_PC},~b,~c) differs from (\ref{tau_cm}) by a change of bracket $(1+h[{m_e}^2/m_{{\nu}_H}^2])$ to
$h[{m_e}^2/m_{{\nu}_H}^2]$.

For $\nu_{H}$, the $e^+e^-$ decay is much faster than the radiative decay. For reference, if $m_{\nu _{H}}$ = 5 MeV
(and $|U_{eH}|^2 \sim $1),
$\tau (\nu_{H}\rightarrow\nu_{L}{e^+e^-}) \approx$ 10 s while $\tau (\nu_{H}\rightarrow\nu_{L}\gamma) \sim$ 10$^{10}$ s. 
\label{eq:decay}

No positive evidence for heavy sterile neutrinos has so far been found in laboratory searches over a wide mass  range. For masses $m_{\nu H} <$ 1 MeV
the most sensitive searches have looked for kinks in the electron spectra of $\beta$-decays and have constrained the sterile-electron mixing to
$|U_{eH}|^2 <10^{-2}$-$10^{-3}$ \cite{Sch83}-\cite{Hol00}. The agreement with expectations of $ft$-values for superallowed pure Fermi
beta-transitions also limits the mixing of massive neutrino and electron flavor \cite{Cal83}-\cite{Deu90}.

The decays of massive antineutrinos from a reactor $\nu_{H}\rightarrow\nu_{L}+e^++e^-$ were looked  for in
\cite{Vog84}-\cite{Hag95}; the latter reference reports on the tightest limit on the mixing parameter, $|U_{eH}|^2
<(0.3-5)\times10^{-3}$, for $m_{\nu_{H}}\sim $(1.1 --9.5) MeV. The limits on  $|U_{eH}|^2$ were obtained in assumption
that only $W$ exchange mode occurs (fig.~\ref{Figure:Decay_PC},b).

Heavy neutrinos with mass up to 15 MeV could be produced in $^8\rm{B}$ decays inside the Sun, and then  decay in
flight. Mixing parameters above {$ |U_{eH}|^2 < (10^{-6}-10^{-4})$ were excluded by measurements of the positron flux
in interplanetary space \cite{Tou81}.

The mixing of heavier neutrinos with both $\nu_e$ and $\nu_{\mu}$ can be tested by looking for peaks  in the leptonic decays of pions and kaons
\cite{Bri92,Aok11}. Alternatively one can look for the products of their decays directly~\cite{Ber88}. The dominant decay channel for a sterile
neutrino of mass 140-500 MeV is $\nu_H\rightarrow \nu +e  +\pi$. For larger neutrino masses, new decay channels open up which include kaons, eta, etc
\cite{Ber86}-\cite{Liv13}. Accelerator experiments with beam of neutrinos from $\pi$- and K- decays constrain the coupling of still heavier neutrinos
(see \cite{Atr09}, \cite{PDG10} and references therein).

Cosmological and astrophysical bounds on sterile neutrino properties are very strong but depend  on the assumptions of
sterile neutrino production. The most stringent bounds on $|U_{eH}|^2$ come from primordial nucleosynthesis and SN1987A
data \cite{Obe93}-\cite{Boy09}. Sterile neutrinos can also have important cosmological implications for big bang
nucleosynthesis and primordial light element abundances, and could distort the spectra of both the cosmic microwave
background and the diffuse extragalactic background radiation \cite{Gel08}.

\begin{figure}
\includegraphics[bb = 104 257 499 600, width=8cm,height=6cm]{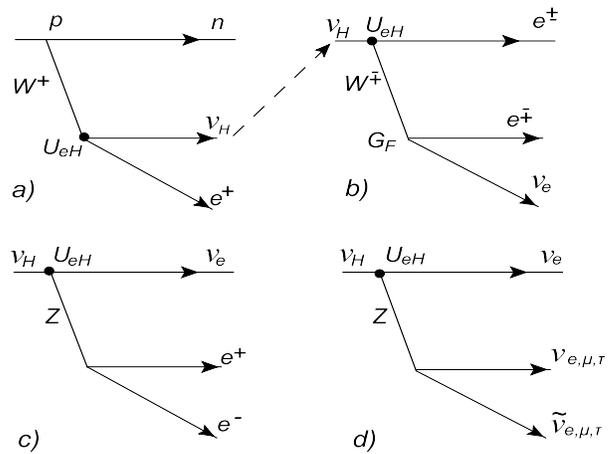}
\caption { Feynman graphs describing the appearance ${\rm{^8B}}\rightarrow {\rm{{^8Be}}} + e^+ + \nu_{H}$ (a) and decay
$\nu_{H}\rightarrow\nu_{e}+e^++e^-$ (b,c) of a heavy neutrino. The invisible decay mode $\nu_{H}\rightarrow\nu_{e}\nu_{i}\widetilde{\nu}_{i}$ is
shown also (d).} \label{Figure:Decay_PC}
\end{figure}

In this paper we present the results of the search for the $\nu_{H}\rightarrow\nu_{L}+e^++e^-$ decay inside the active volume of the Borexino
detector (fig.~\ref{Figure:Decay_PC},~b,~c). For the case of the Dirac neutrinos, only diagram with $(\nu_H e^- W^+)$ vertex contribute to the
$\nu_H$ decay. Two diagrams correspond to the decays of the Majorana neutrinos (fig. \ref{Figure:Decay_PC}, b). The Majorana neutrino total decay
rate is given by (\ref{tau_cm}) multiplied by a factor of 2, which accounts for charge-conjugated decay modes.

We have previously published a search for heavy neutrino decays using the Borexino counting test facility (CTF) \cite{Bac03}.

\section{Experimental set-up and results of measurements}
\subsection {Brief description of Borexino}

The Borexino experiment at the underground Gran Sasso National Laboratory (LNGS) detects solar neutrinos by means  of
their elastic scattering on electrons in a large volume liquid scintillator detector. Thanks to its excellent
radiopurity and large target mass, Borexino is ideal to answer other fundamental questions and look for exotic
processes in particle physics and astrophysics.

The Borexino detector and its components have been thoroughly described in \cite{Ali02}-\cite{Bac12}. The central
detector  comprise of 278 tons of purified organic liquid scintillator confined to a 4.25 m radius transparent nylon
vessel (IV). The IV is  surrounded by a concentric spherical quenched pseudocumene buffer which shield the scintillator
from $\gamma$-rays and neutrinos from the periphery of the detector. The scintillator and buffer are contained in a
13.7 m diameter Stainless Steel Sphere (SSS) on which 2212 photomultipliers (PMTs) are mounted to detect scintillation
from events inside the IV. The SSS is immersed in a water-\v{C}erenkov muon detector.

In Borexino, charged particles are detected through the scintillation light they produce in the liquid scintillator. The energy of an ionizing event
occurring in the scintillator is converted to scintillation light and is quantified by the total light collected by the PMTs. To good approximation,
the measured light depends linearly on the energy released in the scintillator and the energy resolution is scaled as $5\%/\sqrt{E[\rm{MeV}]}$.

The detector energy and spatial resolution were studied with radioactive sources placed at different positions inside
the  inner vessel \cite{Bac12}. For energies $>$3 MeV of interest for this work, the energy calibration was performed
with an $^{241}$Am-$^9$Be neutron source \cite{Bel10,Bel10A}. The position of an event is determined using a photon
time of flight reconstruction algorithm. The position resolution measured using the $^{214}$Bi-$^{214}$Po
$\beta-\alpha$ decay sequence, is 13 cm \cite{Ali09}.

\subsection{Data selection}

The measured Borexino energy spectrum in the 0-15 MeV range from 1192.0 live-days of data and with different selection
cuts  is shown in fig.\ref{Figure:Spectra}. Below 3~MeV the spectrum is dominated by 2.6 MeV $\gamma$'s from
$\beta$-decay of trace $^{208}$Tl in the PMTs and in the SSS.
\begin{figure}
\includegraphics[bb = 30 90 500 760, width=8cm,height=10cm]{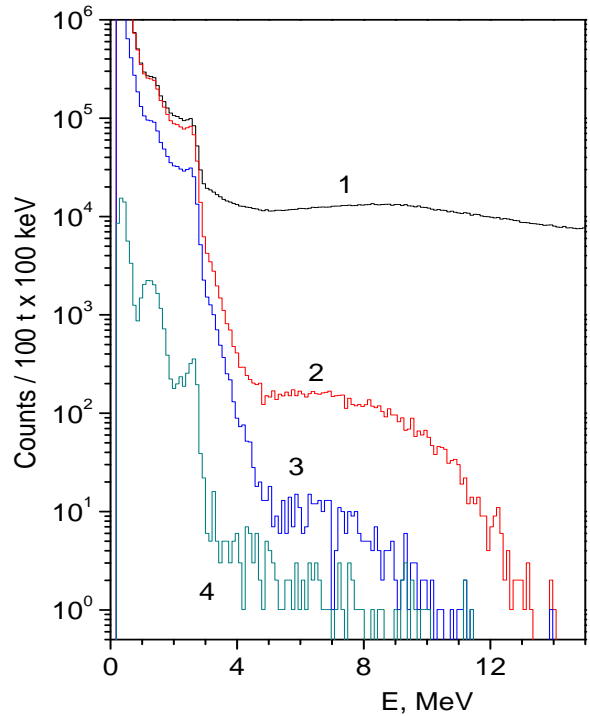}
\caption { Energy spectra of the events surviving incremental selection cuts. From top to bottom: (1) raw spectrum; (2) 2 ms post-muon veto cut; (3)
20 s after muons crossing the SSS cut; (4) FV cut.  See text for details. } \label{Figure:Spectra}
\end{figure}

The first cut on the raw data filters out events occurring within 2 ms of muons crossing the entire detector  (curve 2,
Fig.\ref{Figure:Spectra}). Muons are identified by the outer detector, by the specific mean time of PMT hits when they
cross the SSS and on the time corresponding to the maximum density of hit PMTs.  This timing cut rejects residual muons
that were not tagged by the muon detector and that interacted in the pseudocumene buffer region \cite{Bel11B}.

To remove muon-induced background due to short-lived  isotopes (1.1 s $^8\rm{B}$, 1.2 s $^8\rm{Li}$, etc \cite{Bel10A})
and significantly reduce that from $^{11}\rm{Be}$ ($\tau = 19.9$ s), an additional 20 s veto is applied after each muon
crossing the SSS (curve 3, fig.\ref{Figure:Spectra}).  This cut has a dead time of 745.8 days and brings the live-time
down to 446.2 days.

A software fiducial volume cut is needed in order to suppress external radiation background. Curve 4  of
fig.\ref{Figure:Spectra} shows the effect of selecting a central 100 ton fiducial volume (FV) by applying a radial cut
R $\leq$ 3.02 m. Finally, an $\alpha-\beta$ selection cut based on pulse shape-discrimination performed with the Gatti
optimal filter \cite{Gat62} is applied.  Only events with negative Gatti variable (corresponding to $\gamma$- and
$\beta$-like signals) are selected \cite{Ali09}. Since the energy of $\alpha$ particles is highly quenched in liquid
scintillator, this cut has no effect on the spectrum for energies higher than 4 MeV.

\section{Neutrino fluxes and the $e^+e^-$-spectra}

Here we report on the search for heavy neutrinos produced in ${^8\rm{B}}$ decays in one of the side branches of  the
$pp$ fusion reaction chain in the Sun.  The decay, ${^8\rm{B}} \rightarrow {^8\rm{Be}} + e^+ +\nu_H$, is a variant of
the standard decay with a left-handed light neutrino. In Borexino the search is performed by comparing the measured
energy spectrum with that expected from $\nu_H$-decays. The latter requires the knowledge of the heavy neutrino flux
$\Phi (E_{\nu})$ through the detector, of the kinetic energy of the $e^+e^-$ pairs produced (eq.~\ref{eq:decay}), and
the response function of Borexino to energy released by $e^+e^-$ pairs in the scintillator.

The emission of a heavy neutrino in the ${\beta}^+$-decay of $^8\rm{B}$ is suppressed by the mixing parameter $|U_{eH}|^2$ and a
phase-space factor as:
\begin{equation}
  \Phi(E_{{\nu}_{H}})=|U_{eH}|^2 \sqrt{1-\left(\frac{m_{\nu_{H}}}{E_{{\nu}_H}}\right)^2} \Phi_{^8\rm{B}}(E_\nu)
\end{equation}
where $E_{{\nu}_H}$ is the total energy of the heavy neutrino. We use the neutrino spectrum from $^8\rm{B}$
decay $\Phi_{^8\rm{B}}(E_\nu)$ given in \cite{Bah06}-\cite{Ser09}.

Heavy neutrinos produced in the Sun can decay on their flight to Earth. The energy spectrum of neutrinos reaching the
detector is given by
\begin{equation}
  \Phi(E_{\nu})=\exp(-{\tau_{f}}/{\tau_{c.m.}})\Phi_{m_{\nu H}}(E_{{\nu}_H})
\end{equation}
where $1/\tau_{c.m.}=\Gamma_{c.m.}$ defined by (eq.~\ref{tau_cm}) and $\tau_{f}$ is the time of flight in the center
of mass system (c.m.s):

\begin{equation}
 \tau_{f}= \frac{m_{\nu_{H}}}{E_{\nu}}\frac{D}{\beta c}
 \label{tau_fl}
\end{equation}
$D=1.5\times10^{13}$ cm is the average Sun-Earth distance and $\beta=p_{{\nu}_H}/E_{{\nu}_H}$.

The double differential distribution for energy $\epsilon$ and emission angle $\theta$ for the standard decay with a light
neutrino $\nu_L$ for the c.m.s. is given in~\cite{Shr81}:
\begin{equation}
\frac{dN_{\nu_{L}}}{d\epsilon\: d\cos\theta}=\Gamma_{c.m.}(f_{1} + \xi |\overrightarrow{P}| f_{s}\cos\theta) \label{angle_distr}
\end{equation}
where $f_1(\epsilon,m_{\nu_{H}})$ and $f_s(\epsilon,m_{\nu_{H}})$ are complex functions~\cite{Shr81}, $\xi$=+1(-1) for
$\nu_{H}$ ($\widetilde{\nu}_{H}$), and $|\overrightarrow{P}|=\beta$ is the polarization of the $\nu_{H}$.

The distribution (\ref{angle_distr}) holds for Dirac neutrinos. The angular distribution for the decays of the Majorana
neutrino is distinguishable from that of the Dirac neutrino \cite{Lie82}. However, the integrated decay rate remains
the same in both cases \cite{Boe87}.

The total laboratory energy of the $e^+e^-$-pair, $E=E_{\nu_H}-E_{\nu_{L}}$ and $\epsilon$ relate as follows:
\begin{equation}
E = E_{{\nu}_H}(1-(\epsilon/m_{{\nu}_H})(1+\beta\cos\theta))
\end{equation}
In the  c.m.s. the energy $\epsilon$ of the emitted neutrino is restricted according to
$\epsilon$$\leq$$\epsilon_{max}= (m_{\nu_{H}} ^2-4m_e^2)/2m_{\nu_{H}}$ that corresponds to the emission angle

\begin{equation}
(\cos\theta)_{min}=(1/{\beta})((m_{\nu_H}/\epsilon_{max})(1-E/{E_{{\nu}_H}})-1)
\end{equation}
\begin{figure}
\includegraphics[width=9cm,height=10.5cm]{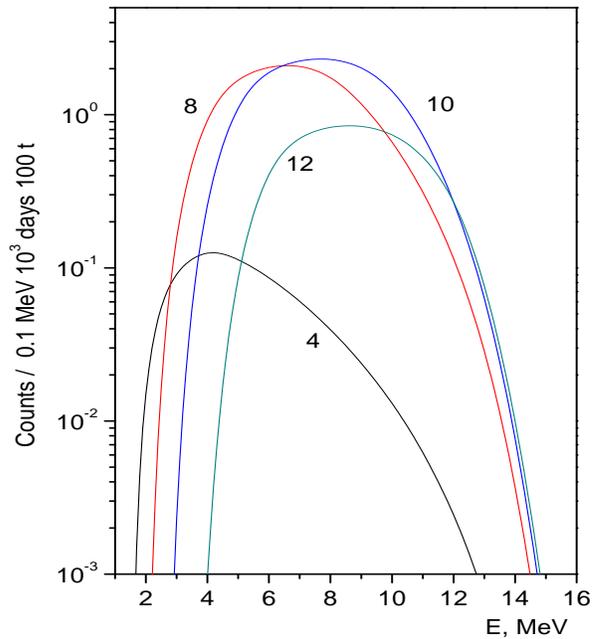}
\caption { The expected spectra of signals due to $\nu_{H}\rightarrow\nu_{L}+e^++e^-$ decay for different neutrino
masses $m_{\nu_{H}}$= 4, 8, 10, 12 MeV and mixing parameter $|U_{eH}|^{2}=1.0\times10^{-5}$.} \label{Figure:Responses}
\end{figure}

The differential spectrum of the $e^+e^-$-pair is obtained by integration of (6) over $\cos(\theta)$ (or $\epsilon$)
and taking into account eq.(7):
\begin{equation}
  \frac{dN}{dE}(E,E_{{\nu}_H})=\int\limits_{(\cos\theta)_{min}}^{1}\frac{dN_{\nu_{L}}}{d\epsilon\: d\cos\theta}d\epsilon d\cos\theta
\end{equation}
For a given heavy neutrino energy $E_{{\nu}_H}$, the energy $E$ of the $e^+e^-$-pair is restricted to the interval
\begin{equation}
  E_{{\nu}_H}[1-\frac{1+\beta}{2}(1-\frac{4m_{e}^{2}}{m_{\nu_{H}}^{2}})]\leq E\leq E_{{\nu}_H}
\end{equation}
Integrating over the neutrino energy up to end-point energy $Q_0$ yields the energy spectrum of the $e^+e^-$ pair:
\begin{equation}
  \frac{dN}{dE}(E)=\frac{(\Gamma_{c.m.}^{e^+e^-})^2}{\Gamma_{c.m.}^{tot}}|U_{eH}|^{2}\frac{m_{{\nu}_H}}{c}\int\limits_{2m_e}^{Q_0}\frac{dN}{dE}(E,E_{\nu})\frac{\Phi(E_{\nu})}{\beta
 E_{\nu}}dE_{\nu}.
    \label{dN/dE}
\end{equation}

Here ${\Gamma_{c.m.}^{e^+e^-}}/{\Gamma_{c.m.}^{tot}}$ is the branching ratio for $\nu\rightarrow\nu e^+e^-$ decay mode. Equation~\ref{dN/dE} is
convolved with the response function of two $e^+e^-$ annihilation $\gamma$ rays and the energy resolution function $\sigma(E)$ to obtain the heavy
neutrino energy spectrum $N^{\nu_H}(E)$. The Borexino response to $e^+e^-$ pairs of different energies was simulated via MC  using GEANT4 and taking
into account the effect of ionization quenching and the non-linear position-dependent energy response. Uniformly distributed events were simulated
inside the entire inner vessel, and those reconstructed within the FV (central 100 t) were used in determining the response function and the
detection efficiency. The expected spectra for different values of heavy neutrino mass $m_{\nu H}$ and for fixed $|U_{eH}|^2=1\times10^{-5}$ are
shown in fig.\ref{Figure:Responses}.


\begin{figure}
\includegraphics[width=9cm,height=10.5cm]{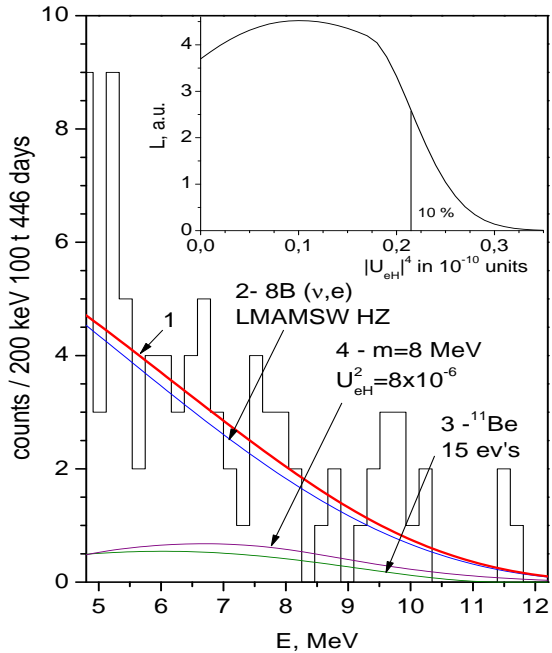}
\caption { The Borexino spectrum in the ($4.8-12.8$) MeV range. 1 - optimal fit;   2 - $(\nu,e)$ - scattering of
$^8\rm{B}$-neutrino (LMA MSW BPS06(GS98); 3- 15 $^{11}\rm{Be}$-decays, 4- the detector response function for the decay
of neutrino with $m_{\nu H}$= 8 MeV and $|U_{eH}|^2 = 8\times 10^{-6}$. The inset shows the dependence of $L$ on
$|U_{eH}|^4$ for $m_{\nu H}$= 8 MeV.} \label{Figure:Fit_region}
\end{figure}

\section{Results and discussions}
\subsection{Fitting procedure}

Figure \ref{Figure:Fit_region}  shows the observed Borexino energy spectrum in the ($4.8-12.8$) MeV range, most
sensitive to the neutrino decay signal. The spectrum includes the $(\nu,e)$ elastic scattering signal from $^8\rm{B}$
solar neutrinos ($N^{^8\rm{B}}(E)$), the rate of long-lived cosmogenic $^{11}\rm{Be}$ ($N^{^{11}\rm{Be}}(E)$) and the
sought-after spectra from neutrino decays ($N^{\nu_H}(E)$):
\begin{equation}
N^{\rm th}(E) = X^{^8\rm{B}}N^{^8\rm{B}}(E) + X^{^{11}\rm{Be}}N^{^{11}\rm{Be}}(E) + X^{\nu_H}N^{\nu_H}(E),\label{Function:fit}
\end{equation}
where $X^{^8\rm{B}}$, $X^{^{11}\rm{Be}}$ and $X^{\nu_H}$ give the intensities of the corresponding processes.

A binned maximum likelihood fit using the expression above yields the number of $(\nu_H\rightarrow \nu_Le^+e^-)$
events.  The likelihood function assumes the form of a product of Poisson probabilities:
\begin{equation}
 L=\prod e^{-N^{\rm th}_i}({N^{\rm th}_i})^{N^{\rm exp}_{i}} / N^{\rm exp}_{i}!\label{Function:Poisson}
\end{equation}
where $N^{\rm th}_i$ and $N^{\rm exp}_i$ are the expected (\ref{Function:fit}) and measured number of counts in the i-th bin of the spectrum,
respectively. The parameter ($X^{^8\rm{B}}$) was constrained by the interval of standard solar model predictions for high and low solar metallicity
\cite{Bah06}-\cite{Ser09}, while ($X^{^{11}Be}$) and $X^{\nu_H}$ were left free. The fit was performed in  the range of 4.8–-12.8 MeV and had 76
degrees of freedom.

The maximum likelihood fit for $m_\nu$=8 MeV is shown in fig.\ref{Figure:Fit_region}, line 1. The best fit corresponds
to the $|U_{eH}|^2$ = $3.7\times 10^{-6}$, $X^{^{11}\rm{Be}}$ = 0 and  $X^{^8\rm{B}}$ for BPSO5(GS98) model
\cite{Bah06}-\cite{Ser09}. The modified $\chi^2 = \sum (N^{\rm exp}_i-N^{\rm th}_i)^2/N^{\rm th}_i$ is 70.5/76. The low
statistics dictate the use of a Monte Carlo simulation of eq.~\ref{Function:fit} to find the probability of $\chi^2_p
\geq$ 70.5. The goodness-of-fit ($ p = 60.5\%$) shows that the function in eq.~\ref{Function:fit} describes the
spectrum well.

\subsection{The obtained limits}

No statistically significant deviations of $|U_{eH}|^2$ from zero were observed for all tested $m_{\nu H}$. The upper limit on the number of
$(\nu_H,\nu_Le^+e^-)$ events for different $m_{\nu_H}$ was found using the $L_{\rm{max}}(X^{\nu_H})$ profile, where $L_{\rm{max}}(X^{\nu_H})$ is the
maximal value of $L$ for fixed $X^{\nu_H}$ when the two other parameters are left free. The $L_{\rm{max}}(X^{\nu_H})$ distribution obtained from MC
simulations with $X^{\nu_H} \geq 0$ was used to determine confidence levels in $L_{\rm max}(X^{\nu_H})$. The upper limits on the mixing parameter
$|U_{eH}|^2$ were then calculated in accordance with Eq.~\ref{dN/dE}.

The dependence of the number of  $(\nu_H,\nu_Le^+e^-)$ counts on $|U_{eH}|^2$ and $m_{\nu H}$ is:
\begin{equation}
 S_{int}(m_{\nu H},|U_{eH}|) \sim m_{\nu H}^6|U_{eH}|^4\exp{({\rm{-const}}\times m_{\nu H}^6|U_{eH}|^2)}
\end{equation}

The function $S_{int}(m_{\nu H},|U_{eH}|)$ has a maximum, for fixed $m_{\nu H}$ and $E_{\nu}$, corresponding  to $|U_{eH}|^2 =2/({\rm{const}}\times
m^6_{\nu_H})= 2(|U_{eH}|^2 \tau_{c.m.})/\tau_f$, where $\tau_{c.m.}=1/\Gamma_{c.m.}$ and $\tau_f$ are defined in equations (\ref{tau_cm}) and
(\ref{tau_fl}). The experiment is not sensitive to low $|U_{eH}|^2$ (due to the low probability of $\nu_H$ decay) nor to large $|U_{eH}|^2$ for which
the $\nu_H$ decays occur in flight from the Sun.

\begin{figure}
\includegraphics[width=9cm,height=10.5cm]{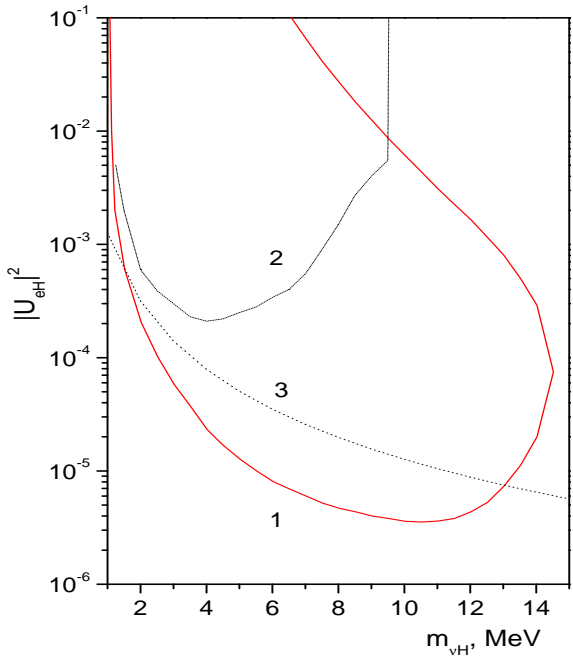}
\caption { Limits on the mixing parameter $|U_{eH}|^{2}$  as a function of neutrino mass $m_\nu$ (90\% c.l.). 1: Present work excludes values of
$|U_{eH}|^{2}$ and $m_\nu$ inside region 1. 2: Upper limits from reactor experiments on the search for $\nu_{H}\rightarrow\nu_{L}+e^++e^-$ decay due
to W-exchange mode \cite{Hag95}. 3: Upper limits from $\pi \rightarrow e + \nu$ decay \cite{Bri92}.} \label{Figure:Limits}
\end{figure}

The Borexino sensitivity curves for $|U_{eH}|^2$ in the $1 \leq m_{\nu H} \leq 15$ MeV range are shown in
fig.\ref{Figure:Limits} together with those obtained by reactor \cite{Hag95} and $\pi^+ \rightarrow e^+\nu_{e}$ decay
experiments \cite{Bri92}.
For the 1.5-13 MeV heavy neutrino mass region Borexino improves the limits on the mixing parameter with
respect to those obtained previously at nuclear reactors and accelerators.

The existing bounds on $|U_{eH}|^2$  in the wide mass range (0.01 - 100) MeV are shown in fig.\ref{Figure:AllLimits}.
Reactor experiments restrict the mixing parameter in the (1.1--9.5) MeV mass range (lines 2, \cite{Der93} and 3,
\cite{Hag95}).  As noted above these limits were obtained in assumption that only $W$ exchange mode can occurs.

For masses up to $m_{\nu H} \approx $ 6 MeV, an important bound is provided by $ft$-values for superallowed Fermi
decays (line 4, \cite{Deu90}) and by searches of kinks in the electron spectrum of $\beta $-decays of ${^{20}\rm{F}}$
(line 5, \cite{Deu90}), ${^{64}\rm{Cu}}$ (line 6, \cite{Sch83}), ${^{45}\rm{Ca}}$ (line 7, \cite{Der97}),
${^{63}\rm{Ni}}$ (line 8, \cite{Hol00}) and ${^{35}\rm{S}}$ (line 9, \cite{Hol00}).

\begin{figure}
\includegraphics[width=9cm,height=10.5cm]{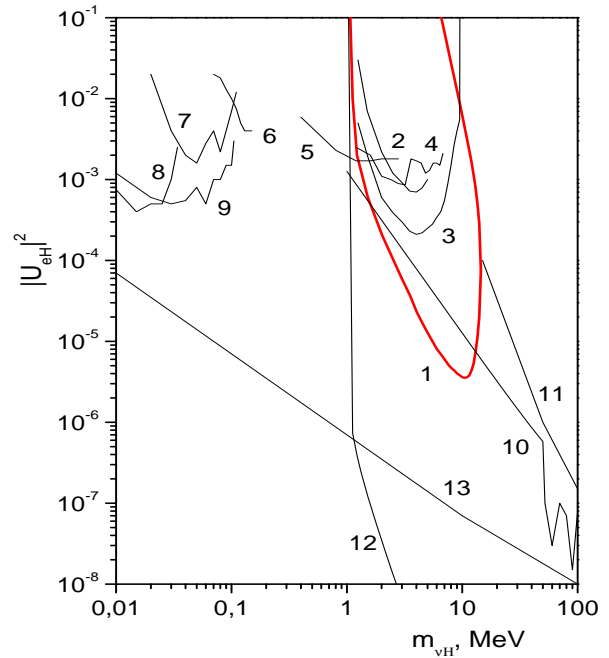}
\caption { The Borexino constraints (region 1) and limits on $|U_{eH}|^2$ versus $m_{\nu H}$ in the mass range
(0.01–-100) MeV from different experiments (see text). }\label{Figure:AllLimits}
\end{figure}

For higher masses, very robust bounds can be set by looking for additional peaks in the spectrum of electrons in
leptonic decays of pions and kaons (line 10, \cite{Bri92}). The data from the experiment PS191 sets a bound which is
strongly mass dependent, going from $|U_{eH}|^2\leq 10^{-4}$ at $m_{\nu H}$ = 15 MeV to $|U_{eH}|^2 \leq
1.5\times10^{-7}$ at $m_{\nu H}$ = 100 MeV (line 11, \cite{Ber88}).

The most restrictive constraints - $|U_{eH}|^2 m_{\nu H}^4 \leq 4.8\times 10^{-7}$ - for neutrino masses up to around
30 MeV were obtained  from the absence of a $\gamma$-ray signature in conjunction with the neutrino burst from SN1987A.
(line 12, \cite{Obe93}).

For $m_{\nu H}$  in the range (100 MeV - 100 GeV) the limits $|U_{eH}|^2 \leq (10^{-8}-10^{-3})$ were obtained in NA3
\cite{Bad86}, CHARM \cite{Ber86}, L3 \cite{Adr92}, DELPHI \cite{Abr97} and Belle \cite{Liv13} experiments. Similar
limits for sterile neutrino mixing with muon neutrino $U_{\mu H}$ can be found in \cite{Atr09}.

Finally, if sterile neutrinos are Majorana particles, they would contribute to the mediation of neutrinoless double beta decay. The limit on the
half-life time of this process can be translated into a bound on the mixing parameter $|U_{eH}|^2$, which scales as $m_{\nu H}$ for $m_{\nu H} \leq$
30 MeV and as $m_{\nu H}^{-1}$ for $m_{\nu H}\geq$ 400 MeV, with $|U_{eH}|^2\approx 10^{-8}$ at $m_{\nu H}$ = 100 MeV (line 13, \cite{Ben05, Sem10,
Mit12}).

\section{Conclusion}
Thanks to the uniquely low radioactive background and large target mass of the Borexino detector, new limits on the
mixing parameter $|U_{eH}|^2$  of a hypothetical massive neutrino in the range of mass 1.5 - 14 MeV to the electron
neutrino have  been set. These limits are 10- to 1000-fold stronger than those obtained by experiments searching for
$\nu_{H}\rightarrow\nu_{L}+e^++e^-$ decays at nuclear reactors and 1.5-4 times stronger than those inferred from
$\pi\rightarrow e + \nu$ decays.

\section{Acknowledgements}
The Borexino program was made possible by funding from INFN and PRIN 2007 MIUR (Italy), NSF (USA), BMBF, DFG, and MPG
(Germany), NRC Kurchatov Institute (Russia), Polish National Science Center (grant DEC-2012/06/M/ST2/00426) and RFBR
(grants 13-02-01199, 13-02-12140-ofi-m and 13-02-92440-ASPERA). We acknowledge the generous support of the Laboratori
Nazionali del Gran Sasso (LNGS). A.~Derbin, L.~Ludhova, V.~Muratova and O.~Smirnov acknowledge the support of
Fondazione Cariplo.


\newpage
\begin{thebibliography}{99}

\bibitem{Shr81}R.E.~Shrock, Phys.~Rev. D24, 1232, 1275 (1981)
\bibitem{Lie82} L.F.~Li and F.~Wilczek, Phys.~Rev. D25, 143 (1982)
\bibitem{Pal82} P.B.~Pal and L.~Wolfenstein, Phys.~Rev. D25, 766 (1982)
\bibitem{Boe87}F.~Boem, P.~Vogel, Physics of massive neutrinos. Cambridge University Press, (1987)
\bibitem{Moh91} R.N.~Mohapatra and P.B.~Pal, Massive Neutrinos in Physics and Astrophysics, World Scientific, (1991)

\bibitem{Gor07} D.~Gorbunov, M.~Shaposhnikov, JHEP 0710:015, (2007)
\bibitem{Gni13} S.N.~Gninenko, D.S.~Gorbunov, M.E.~Shaposhnikov, arXiv:1301.5516
\bibitem{Ber03} V.~Berezinsky, M.~Narayan, F.~Vissani, Nucl.~Phys. B658, 254 (2003)
\bibitem{Chu96} E.J.~Chun, A.S.~Joshipura, A. Yu. Smirnov, Phys. Rev. D54, 4654 (1996)
\bibitem{Ben97} K.~Benakli and A.Yu.~Smirnov, Phys.~Rev.~Lett. 79, 4314 (1997)
\bibitem{Cir05} M.~Cirelli, G.~Marandella, A.~Strumia and F.~Vissani, Nucl.~Phys. B708, 215 (2005)
\bibitem{Moh06} R.N.~Mohapatra, A.Y.~Smirnov, Ann.~Rev.~Nucl.~Part.~Sci. 56, 569 (2006)
\bibitem{Hol02} P.C.~de~Holanda, A.Yu.~Smirnov, hep-ph/0211264

\bibitem{Sch83}K.~Schreckenbach, G.~Colvin, F.~von Feilitzsch, Phys.Lett. B129, 265 (1983) 
\bibitem{Sch96} S.~Sch$\ddot{\rm{o}}$nert, L.~Oberauer, C.~Hagner et al., Nucl.~Phys. B48, 201 (1996) 
\bibitem{Der97} A.V.~Derbin et al., JETP Lett. 66, 88 (1997)  
\bibitem{Hol00} E.~Holzschuh et al. Phys.~Lett. B482, 1 (2000)  
\bibitem{Hol00} E.~Holzschuh et al. Phys.~Lett. B451, 247 (1999)  
\bibitem{Cal83} F.P.~Calaprice, D.J.~Millener, Phys.~Rev. C27, 1175 (1983) 
\bibitem{Isa86} V.I.~Isakov, M.I.~Strikman, Phys.~Lett. B181,195 (1986) 
\bibitem{Deu90} J.~Deutsch, M.~Lebrun, R.~Prieels, Nucl.~Phys. A518, 149 (1990)

\bibitem{Vog84}P.~Vogel, Phys.~Rev. D30, 1505 (1984)
\bibitem{Obe87}L.~Oberauer, F.~Von Feilitzsch and R.L.~Mossbauer, Phys.~Lett.  B198, 113 (1987)
\bibitem{Kop90}V.I.~Kopeikin, L.A.~Mikaelyan and S.A.~Fayans, JETP Lett. 51, 86 (1990)
\bibitem{Der93}A.V.~Derbin, et al., JETP Lett. 57, 768 (1993)  
\bibitem{Hag95}C.~Hagner, et al., Phys.~Rev. D52, 1343 (1995) 

\bibitem{Tou81}D.~Toussaint, F.~Wilczek, Nature 289, 777 (1981) 

\bibitem{Bri92} D.I.~Britton et al., Phys.~Rev. D46, 885 (1992) 
\bibitem{Aok11} M. Aoki et.al. (PIENU Coll.), Phys.~Rev. D84, 052002 (2011).
\bibitem{Ber88} G.~Bernardi et al., Phys.~Lett. B203, 332 (1988) 
\bibitem{Ber86} F.~Bergsma et al., (CHARM Coll.), Phys.~Lett. B166, 473 (1986)
\bibitem{Bad86} J.~Badier et al. (NA3 Coll.), Z.~Phys. C31, 21 (1986)
\bibitem{Adr92} O.~Adriani et al., (L3 Coll.), Phys.~Lett. B295, 371 (1992)
\bibitem{Abr97} P.~Abreu et al., (DELPHI Coll.), Z.~Phys. C74, 57 (1997)
\bibitem{Liv13} D.~Liventsev et al., (Belle Coll.), Phys.~Rev. D87, 071102 (2013)

\bibitem{Atr09} A.~Atre, T.~Han, S.~Pascoli, B.~Zhang, JHEP 05, 030 (2009) 
\bibitem{PDG10} K.~Nakamura et al., (Particle Data Group), J.~Phys. G37, 075021 (2010)

\bibitem{Obe93} L.~Oberauer, C.~Hagner, G.~Raffelt and E.~Rieger, Astropart.~Phys.  1, 377 (1993)
\bibitem{Dol00} A.~D.~Dolgov, S.H.~Hansen, G.~Raffelt and D.V.~Semikoz, Nucl. Phys. B580, 331 (2000)
\bibitem{Dol02} A.~D.~Dolgov, Phys.~Rep. 370, 333 (2002)
\bibitem{kus09} A.~Kusenko, Phys.~Rept. 481, 1 (2009)
\bibitem{Boy09} A.~Boyarsky, O.~Ruchayskiy and M.~Shaposhnikov, Ann.~Rev.~Nucl.~Part.~Sci. 59, 191 (2009)
\bibitem{Gel08} G.~Gelmini, E.~Osoba, S.~Palomares-Ruiz, S.~Pascoli, JCAP 0810:029 (2008) arXiv:0803.2735

\bibitem{Bac03} H.O.~Back et al. (Borexino Coll.), JETP Lett., 78, 707 (2003)
\bibitem{Ali02} G.~Alimonti et al. (Borexino Coll.),  Astropart.~Phys. 16, 205 (2002)
\bibitem{Arp08} C.~Arpesella et al. (Borexino Coll.), Phys.~Lett. B568, 101 (2008)
\bibitem{Arp08A} C.~Arpesella et al. (Borexino Coll.), Phys.~Rev.~Lett. 101, 091302 (2008)
\bibitem{Ali09} G.~Alimonti et al. (Borexino Coll.), Nucl.~Instr.~and~Meth. A600, 58 (2009)
\bibitem{Bel10} G.~Bellini et al. (Borexino Coll.), Phys.~Rev. C81, 034317 (2010)
\bibitem{Bel10A} G.~Bellini et al. (Borexino Coll.),  Phys.~Rev. D82, 033006 (2010)
\bibitem{Bel11} G.~Bellini et al. (Borexino Coll.), Phys.~Lett. B696, 191 (2011)
\bibitem{Bel11A} G.~Bellini et al. (Borexino Coll.), Phys.~Rev.~Lett. 107, 141302 (2011)
\bibitem {Bel11B} G.~Bellini et al. (Borexino Coll.), JINST, 6, P05005 (2011)
\bibitem{Bel12A} G.~Bellini et al. (Borexino Coll.), Phys.~Rev.~D85, 092003 (2012)
\bibitem{Bel12B} G.~Bellini et al. (Borexino Coll.), Phys.~Lett. B707, 22 (2012)
\bibitem{Bel12C} G.~Bellini et al. (Borexino Coll.),  Phys.~Rev.~Lett. 108, 051302 (2012)
\bibitem{Bac12} H.~Back et al. (Borexino Coll.), arXiv:1207.4816

\bibitem{Gat62} E.~Gatti and F.~De Martini, Nuclear Electronics, IAEA Wien, 2, 265 (1962)


\bibitem{Bah06} J.N.~Bahcall, A.M.~Serenelli, and S.~Basu, Astrophys.~J.~Suppl. 165, 400 (2006)
\bibitem{Pen08} C.~Pena-Garay, A.M.~Serenelli: arXiv:0811.2424 (2008).
\bibitem{Ser09} A.~Serenelli,  Astrophys.~Space Sci. 328, 13 (2010) arXiv:0910.3690 (2009)

\bibitem{Ben05}P.~Benes, A.~Faessler, F.~Simkovic and S.~Kovalenko, Phys.~Rev. D71, 077901 (2005)
\bibitem{Sem10} F.~Simkovic, J.~Vergados, A.~Faessler, Phys.~Rev. D82,  113015 (2010)
\bibitem{Mit12} V.~Mitra, G.~Senjanovic, F.~Vissani, Nucl.~Phys. B856, 26 (2012)


\end{thebibliography}
\end{document}